# Exploration and Analysis of Combinations of Hamming Codes in 32-bit Memories

David Freitas, David Mota, Clailton Lopes, Daniel Simões, Jarbas Silveira, João Mota and César Marcon

***Abstract*—Reducing the threshold voltage of electronic devices increases their sensitivity to electromagnetic radiation dramatically, increasing the probability of changing the memory cells' content. Designers mitigate failures using techniques such as Error Correction Codes (ECCs) to maintain information integrity. Although there are several studies of ECC usage in spatial application memories, there is still no consensus in choosing the type of ECC as well as its organization in memory. This work analyzes some configurations of the Hamming codes applied to 32-bit memories in order to use these memories in spatial applications. This work proposes the use of three types of Hamming codes: Ham(31,26), Ham(15,11), and Ham(7,4), as well as combinations of these codes. We employed 36 error patterns, ranging from one to four bit-flips, to analyze these codes. The experimental results show that the Ham(31,26) configuration, containing five bits of redundancy, obtained the highest rate of simple error correction, almost 97%, with double, triple, and quadruple error correction rates being 78.7%, 63.4%, and 31.4%, respectively. While an ECC configuration encompassed four Ham(7.4), which uses twelve bits of redundancy, only fixes 87.5% of simple errors. Additionally, we performed a relative reliability analysis among the analyzed codes over 3500 days of operation.**

## I. Introduction

The demand for performance enhancement and power dissipation reduction in various electronic devices requires the Integrated Circuit (IC) industry to increase the integration degree of these devices. In turn, this integration forces the reduction of the threshold voltage of the electronic devices, dramatically increasing the sensitivity of these devices to electromagnetic radiation. Thus, both in space and on earth, the content of memory cells, registers, latches, and flip-flops of ICs can be affected by the radioactive events [1]-[5].

The occurrence of these events has been a research issue for over 40 years, as seen in the works [6][7][8]. Radiation can bring many temporary and permanent worries, such as reducing performance, changing data computing, and damaging electronic components. The combined effects of radiation, reduced voltage, and temperature reduce the operating margin of the circuits causing unexpected faults [9]. Consequently, the need to mitigate faults in these systems arises, aiming at avoiding unwanted behaviors [1][2].

We highlight the Triple Modular Redundancy (TMR) among the techniques proposed to mitigate faults. The TMR bases define the use of three identical logical blocks performing the same tasks; a voting circuit compares the outputs of these blocks and chooses the output value using majority voting. Implementing the TMR technique requires a large amount of area, consequently increasing the IC power dissipation. The works [10]-[12] use TMR to avoid faults in memory elements. Sánchez, Entrena, and Kastensmidt [10] employ the Approximate TMR (ATMR) technique, a TMR scheme where the redundant blocks are a simplified version of the basic block to reduce area and power. She and Li [11] implement TMR to reduce the configuration bits until the calculated fault probability is less than the application's required fault rate. Hoque et al. [12] use TMR to analyze the reliability properties of critical security systems of FPGA-based space applications.

The Error Correcting Code (ECC) is another option widely used to mitigate application faults. ECCs capable of correcting simple errors and detecting double errors in a single word are called Single Error Correction, Double Error Detection (SECDED). The works [13] and [14] present SECDED ECCs targeting spatial applications. Castro et al. [14] introduce the CLC algorithm to detect and correct multiple errors in memory devices using extended Hamming and parity bits. Morán et al. [13] propose a low redundancy ECC to correct Multiple Cell Upsets (MCUs) with less area and power costs than that proposed in [14].

Hard error causes permanent failure in a device, whereas soft error is characterized by corrupting only the information while keeping the device operational. This work proposes and analyzes ECCs intending to treat only soft errors. Among the soft errors are the events that change only one bit: Single Bit Upset (SBU) [5][16], and events that change more than one bit: Multiple-Bit Upset (MBU) and Multiple Cell Upset (MCU). While MCU is defined as the event that changes the content of two or more memory cells or latches, MBU is a type of MCU in which the cells changed are in the same word [4][17].

David C. C. Freitas, David F. M. Mota, Clailton A. Lopes, Daniel L. Simões, Jarbas A. N. da Silveira and João C. M. Mota are with Post-Graduated Program on Teleinformatics Engineering, Federal University of Ceará UFC, Fortaleza – CE, Brazil.
César Marcon is with School of Technology on the Pontifical Catholic University of Rio Grande do Sul PUCRS, Porto Alegre – RS, Brazil.

Although there are various studies on using ECCs in spatial application memories, there is still no consensus on the choice of the best SECDED code or the best organization of these codes in memory, as well as on the choice of error patterns used for correction testing. The investigations take into account code sizes and redundancy types, which are analyzed using error patterns generated randomly in memories fabricated in specific technologies, as exemplified in [13][14][18]. The works [14][19]-[24] show that the codes most used in these applications are BCH, Hamming, Matrix, Reed-Muller, LDPC, Golay, Reed-Solomon, and their variations.

The literature presents the study of several error patterns. Castro et al. [14] generate pseudorandom words for each scenario where cells are positioned adjacently. Radiation tests are also used to characterize the format of these patterns. Radaelli et al. [25] obtain error patterns using tests with different energy levels (i.e., 22 MeV, 47 MeV, 95 MeV, and 144 MeV). There are also researches, like [26], that account for error patterns and the probability of error occurrence.

Several techniques propose different code organizations to obtain different tradeoffs for implementation and reliability. Castro et al. [14] propose a code based on the extended Hamming with parity bits. The work of Silva et al. [27] and Guo et al. [28] present codes based on two-dimensional ECCs [29]. Park, Lee, and Roy [30] propose a technique of product code type to build a larger length code with higher correction capacity using smaller size codes. Poolakkaparambil et al. [31] propose a code based on parity and the BCH code, which was proposed in 1959 by Bose, Chaudhuri, and Hocquenghem and whose initials give the name to the code; this code forms a class of cyclic ECCs that are constructed using polynomials on the Galois Field [32].

This work analyzes SECDED code configurations, considering typical fault patterns of spatial applications. We chose Hamming as a SECDED code in the formats Ham(31,26), Ham(15,11), and Ham(7,4) to allow mixed exploitation of logical and physical protection formats in 32-bit memories.

## II. Background and Methodology

R. Hamming [15] has developed the *Ham(n,k)* linear block code for error detection and correction, where *n* is the total bits of the coded word and k is the number of bits of information or data bits. Additionally, Equation 1 shows that r is the number of parity or redundancy check bits added to the data to compose the codeword.

$$r = n - k \tag{1}$$

Fig. 1 shows that Hamming is a binary linear block code with $r \geq 2$, which is based on Equations 2 and 3.

$$n = 2^r - 1 \tag{2}$$

$$k = 2^r - r - 1 \tag{3}$$

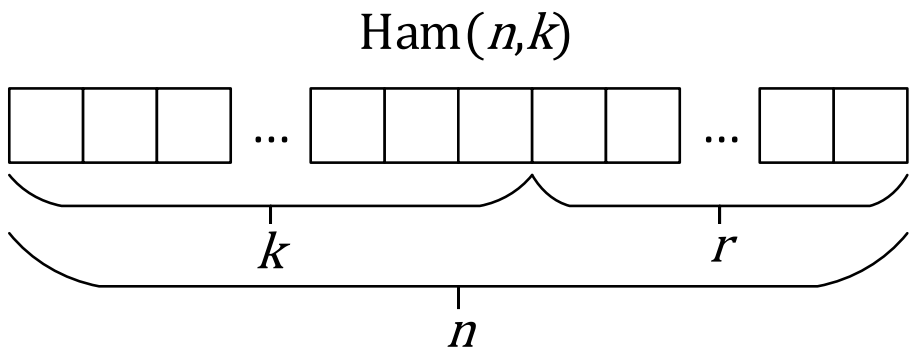

Fig. 1 Representation of a generic Hamming code Ham($n$, $k$); $n$ is the total number of bits, $k$ is the amount of data bits, and $r$ is redundancy.

The Hamming codes used in this work imply the following parameters, Ham(7,4): $n = 7$, $k = 4$, $r = 3$; Ham(15,11): $n = 15$, $k = 11$, $r = 4$; and Ham (31,26): $n = 31$, $k = 26$, $r = 5$. Fig. 2 shows the memory and the configurations of the Hamming codes used in the experiments.

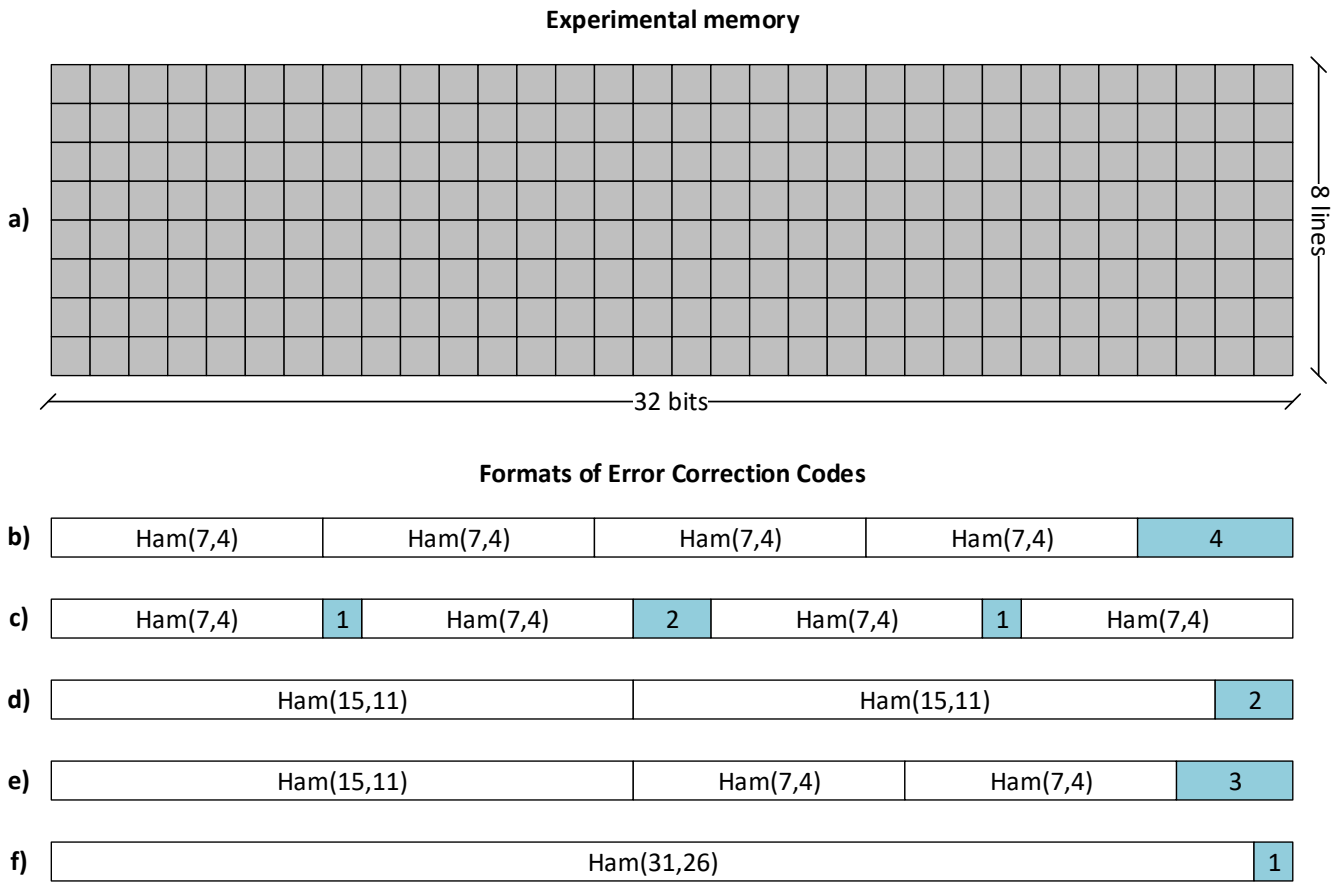

Fig. 2 Experimental memory and Hamming code organizations in the memory lines. a) Illustrates an 8×32 memory, i.e., eight lines of 32 bits. b) Shows the $Ham_{7,4_A}$ configuration, where each of the eight memory lines contains four Ham(7,4), leaving the four final bits uncovered. c) Illustrates $Ham_{7,4_B}$, a configuration similar to $Ham_{7,4_A}$, but the discovered bits are scattered in the word. d) Depicts the $Ham_{15,11}$ configuration encompassing two Ham(15,11) and two bits uncovered. e) Shows that the $Ham_{15,11_7,4}$ configuration combines two Ham(7,4) and one Ham(15,11), leaving three bits uncovered. f) Shows the $Ham_{31,26}$ configuration employs a Ham(31,26), leaving only one bit uncovered.

Fig. 2b to Fig. 2f represent the positions of the Hamming codes in each memory line used in the experiments. For instance, Fig. 2d shows that in each memory line, the $Ham_{15,11_7,4}$ configuration has one Ham(15,11), two Ham (7,4), and three uncovered bits. Consequently, all eight memory lines of this example have 8 Ham(15,11), 16 Ham(7,4), and 24 bits not covered by the ECCs.

Fig. 3 describes the methodology for obtaining the experimental results, containing a flow with the five main activities. Activity ① presents the organization of the Hamming codes in the five configurations described in Fig. 2. Activity ② describes the types of errors generated in the experiment. This step contemplates errors of one, two, three and four bit-flips, performing thirty-six possibilities. Fig. 4 illustrates the error patterns. Activity ③ refers to the positioning of the error patterns from one of the 256 positions of the example, i.e., an eight-line memory containing 32-bit lines. Activity ④ shows that all generated error patterns are verified in all 256 possible positions for the five proposed code configurations. Finally, Activity ⑤ illustrates that all tests are performed through simulation; the simulator output shows how many bits, by error pattern, have been corrected and/or detected.

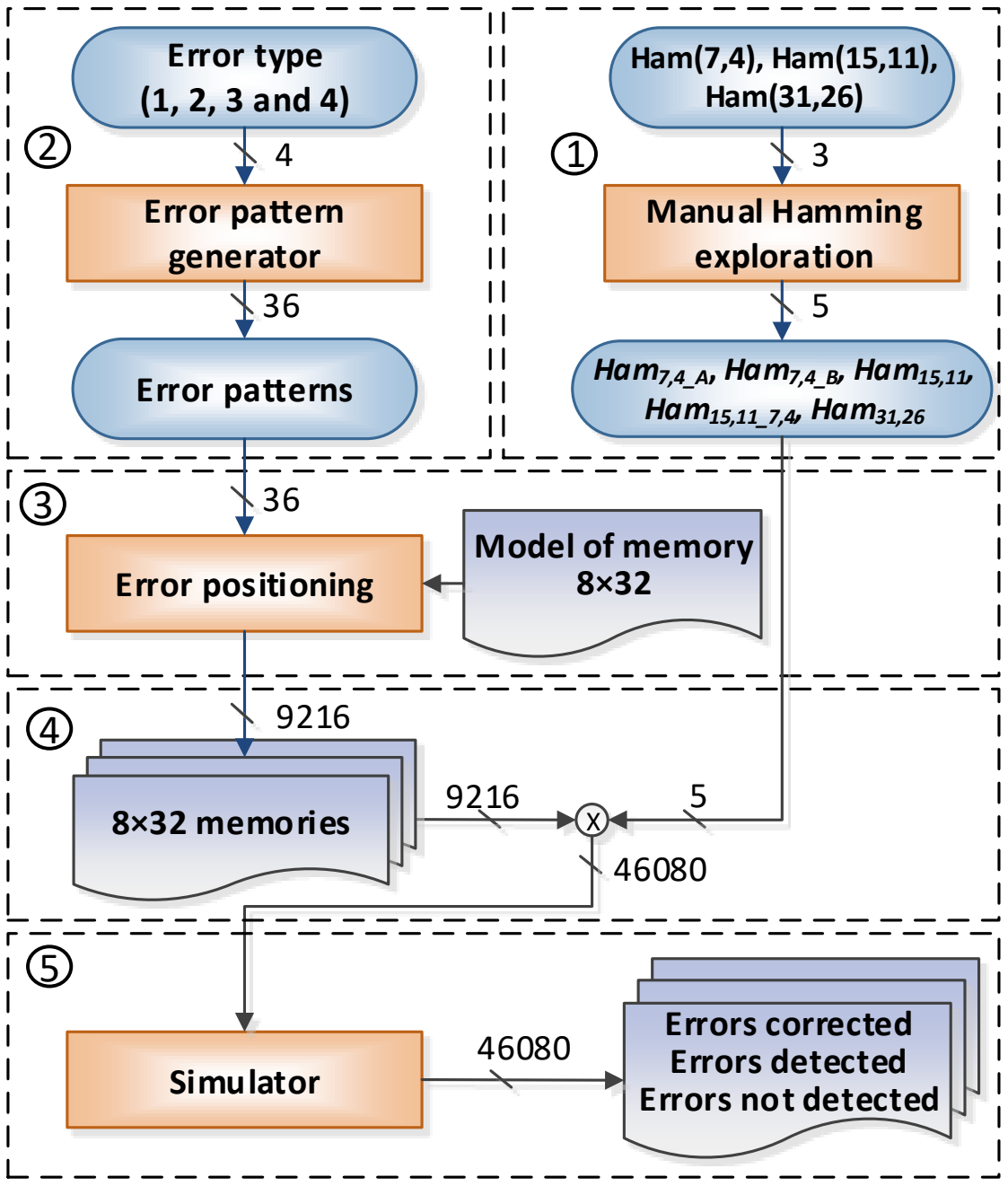

Fig. 3 Methodology applied to the work, containing the five main activities used in the experiments.

Fig. 4 shows the error patterns used in the experiments. These error patterns were based on the work of Rao et al. [26], that performed a neutron particle strike simulation with a commercial evaluation tool.

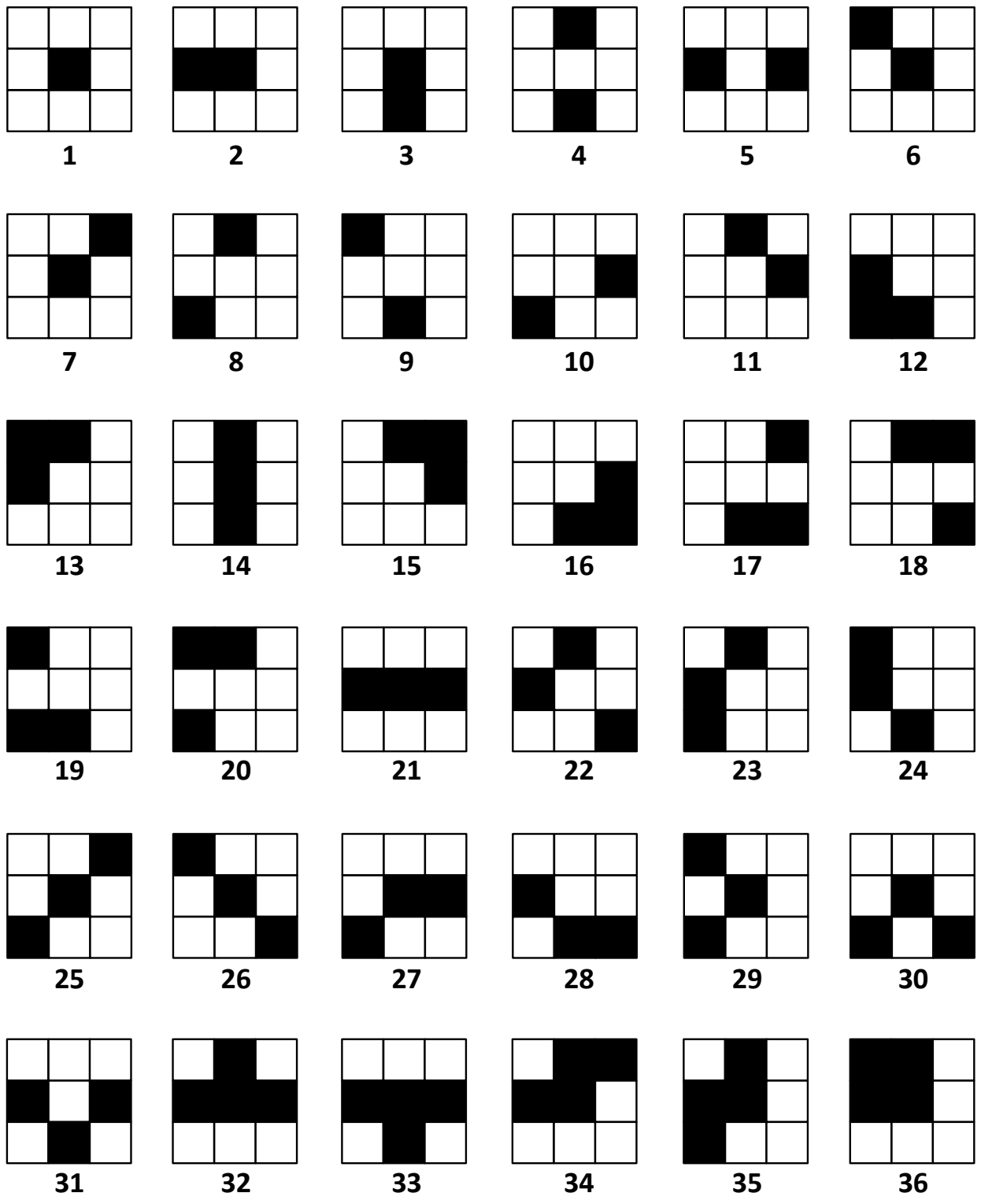

Fig. 4 Thirty-six error patterns used in the experiments, encompassing one simple error, ten double errors, twenty triple errors, and five quadruple errors.

We used computational simulation software to check all 36 error patterns of Fig. 4 in the 256 possible memory cells. Activity ③ of Fig. 3 shows that this task encompasses 9216 possible combinations. We generated the error patterns sequentially. For example, pattern 1 of Fig. 4 was generated in all 256 memory locations. Thus, it was generated in the first position of the 8×32 (1,1) memory (i.e., the leftmost column and uppermost row bit). Then it was generated in position (1,2), next in position (1,3), and so on until the position (8,32), totaling 256 possibilities.

Activities ④ and ⑤ of Fig. 3 show that the five ECC configurations of Fig. 2 were verified for each error pattern generated at each memory location. For example, regarding the $Ham_{7,4_A}$ configuration, the insertion of an error in the memory position (1,1) causes all eight memory lines to be checked by the four Ham(7.4) in the first twenty-eight positions of each line since the ECCs of this configuration do not cover the last four bits of each line.

Near the edges of memory, more specifically in the two leftmost columns and two lowermost rows, the positioning of some error patterns implies that one or more bit-flips can be positioned out of the memory region; for these cases, the error pattern has a smaller size than those shown in the 3×3 matrices of Fig. 4. For example, assuming that the error pattern 7 was positioned on the penultimate line and penultimate column of the memory, i.e., position (7,31) of memory, it will consist of a 2×2 matrix, having only one flip-bit in the memory location (8,32); the other error of pattern 7 is discarded.

Activity ⑤ of Fig. 3 shows that the simulation of each error pattern for each configuration results in three variables: (i) DC (Detected and Corrected) - reporting the number of bits that the Hamming codes detected and corrected; (ii) DNC (Detected, but Not Corrected) - containing the number of bits that were detected but not corrected; and (iii) ND (Not Detected) - informing the number of bit-flips that were not detected by the Hamming codes, and consequently, they are uncorrected errors.

Subsequently, we describe the evaluation of the $Ham_{7,4_A}$ configuration with the error pattern 1 of Fig. 4 to exemplify the use of these variables. When inserting this pattern in the first memory location (1,1), the first code Ham(7,4) detects and corrects the bit-flips; thus, DC = 1, DNC = 0, ND = 0. By inserting this same error pattern into the memory position (1,2), the same situation occurs, causing the DC variable to be incremented. This same result occurs in the first line up to the memory position (1,28), causing the simulation to have DC = 28, DNC = 0, and ND = 0. By inserting the error pattern in position (1,29), an undetected error will occur because the last four bits are not covered by any code Ham(7.4), causing the ND variable to be incremented. At the end of the simulation, the error control variables will be DC = 224, DNC = 0, ND = 32, or in percentage , DC = 87.5%, DNC = 0%, ND = 12.5%.

## III. Results and Discussion

This section presents the simulation results of the five proposed Hamming code configurations, the data presented in three figures highlighting DC, DNC, and ND, in Fig. 5 to Fig. 7, respectively.

Each figure shows the percentage of the DC, DNC, and ND rate for each of the four possibilities of bit-flips that are divided into four scenarios presenting the average values of the occurrence of: (i) only one bit-flip (error pattern 1 of Fig. 4); (ii) two bit-flips (error patterns 2 to 11 of Fig. 4); (iii) three bit-flips (error patterns 12 to 31 of Fig. 4), and (iv) four bit-flips (error patterns 32 to 36 of Fig. 4). These four scenarios are called G1, G2, G3, and G4, respectively.

### *A. Results of Detected and Corrected Errors*

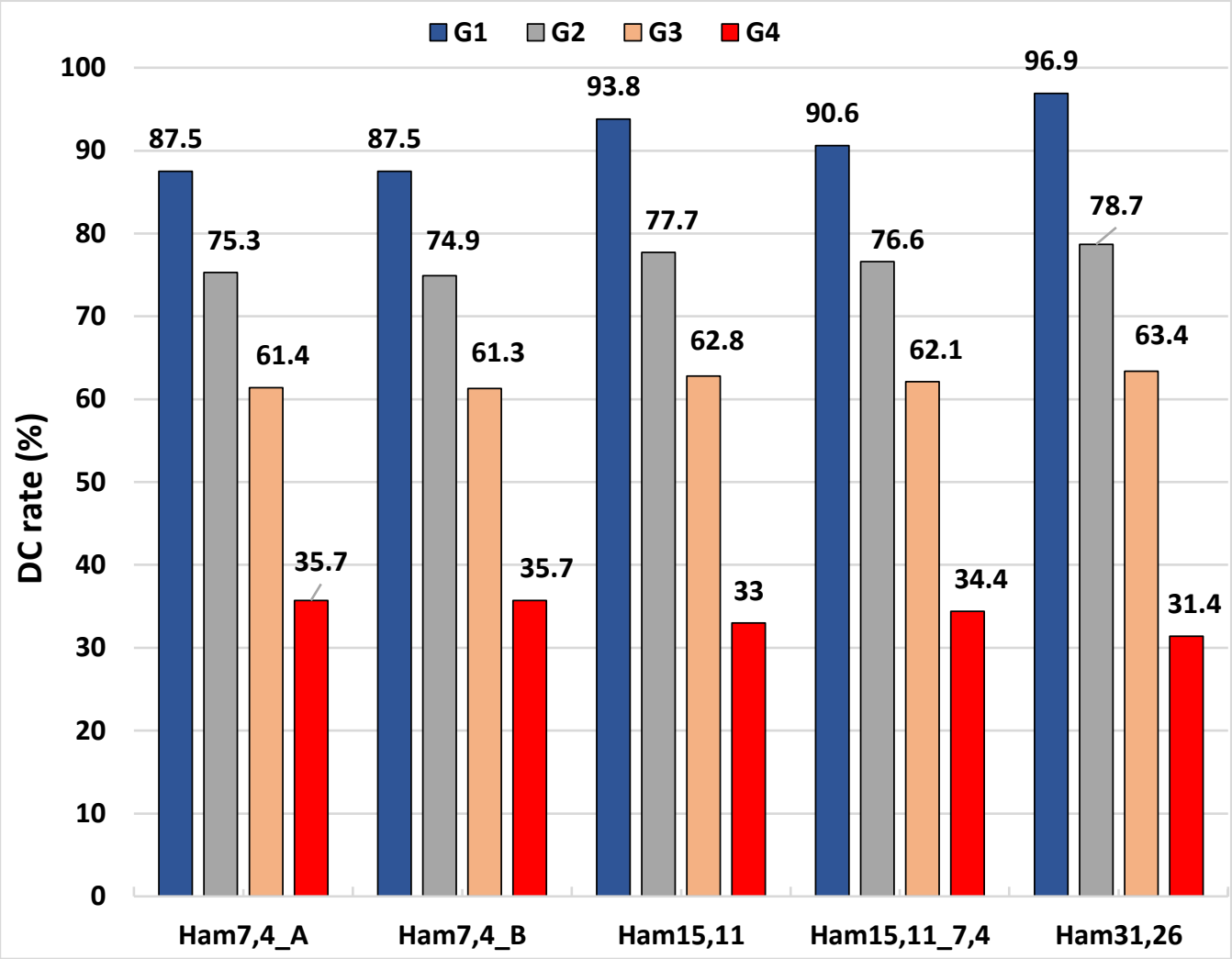

Fig. 5 Simulation results regarding **DC** rate for the five ECC configurations and four bit-flip scenarios: (G1) a single bit-flip; (G2) two bit-flips; (G3) three bit-flips; and (G4) four bit-flips.

Fig. 5 illustrates that the $Ham_{31,26}$ configuration achieved the highest DC rate for the G1 scenario, i.e., 96.9%, followed by the $Ham_{15,11}$ configuration with 93.8%. The configurations employing Ham(7,4) (i.e., $Ham_{7,4_A}$ and $Ham_{7,4_B}$) obtained the lowest DC rates for the G1 scenario, both with 87.5%. $Ham_{31,26}$ reached a high DC rate because this configuration has only a single bit not covered by the ECC. For scenarios G2 and G3, which have more bit-flips, all the configurations achieved very close DC rates; besides, the $Ham_{31,26}$ configuration reached the highest value for both scenarios, 78.7%, and 63.4%, respectively. This proximity in the DC rates suggests that although the configurations with Ham(7,4) have more detection and error correction capacities (at the limit of 4 error corrections per memory line, i.e., a correction for each one of the four Ham(7,4)), the number of uncovered bits and the bit-flip grouping patterns are decisive factors for the reduced DC rate in these configurations. The $Ham_{7,4_A}$ and $Ham_{7,4_B}$ configurations had the highest DC rates for the G4 scenario, which always has error patterns with four bit-flips, both configurations with 35.7%. This is justified because, in many cases, these error patterns generate bit-flips in two nearby Ham(7,4), allowing correcting up to two errors per line.

Finally, Fig. 5 allows us to conclude that as the number of bit-flip increases, the DC rates reduce since all configurations include only SECDED ECCs. On the one hand, although $Ham_{31,26}$ has the least redundancy bits, the configuration obtains the highest DC rates for the G1, G2, and G3 scenarios. These high rates are because $Ham_{31,26}$ is the configuration with fewer bits uncovered, and the bit-flips present in the error patterns are highly grouped. On the other hand, for the G4 scenario, the $Ham_{31,26}$ configuration ranked last. It happens because four bit-flips significantly increase the occurrence of patterns where errors can be corrected by more than one Hamming code positioned in the neighborhood.

### *B. Results of Detected but Not Corrected Errors*

Fig. 6 shows the NDC rates for the five ECC configurations and four scenarios evaluated. All configurations achieved NDC equal to 0 for the G1 scenario because SECDED codes always correct the occurrence of a bit-flip. For the G2, G3, and G4 scenarios, the $Ham_{31,26}$ configuration shows the highest detection rate with 17.9%, 32.3%, and 62.1%, respectively.

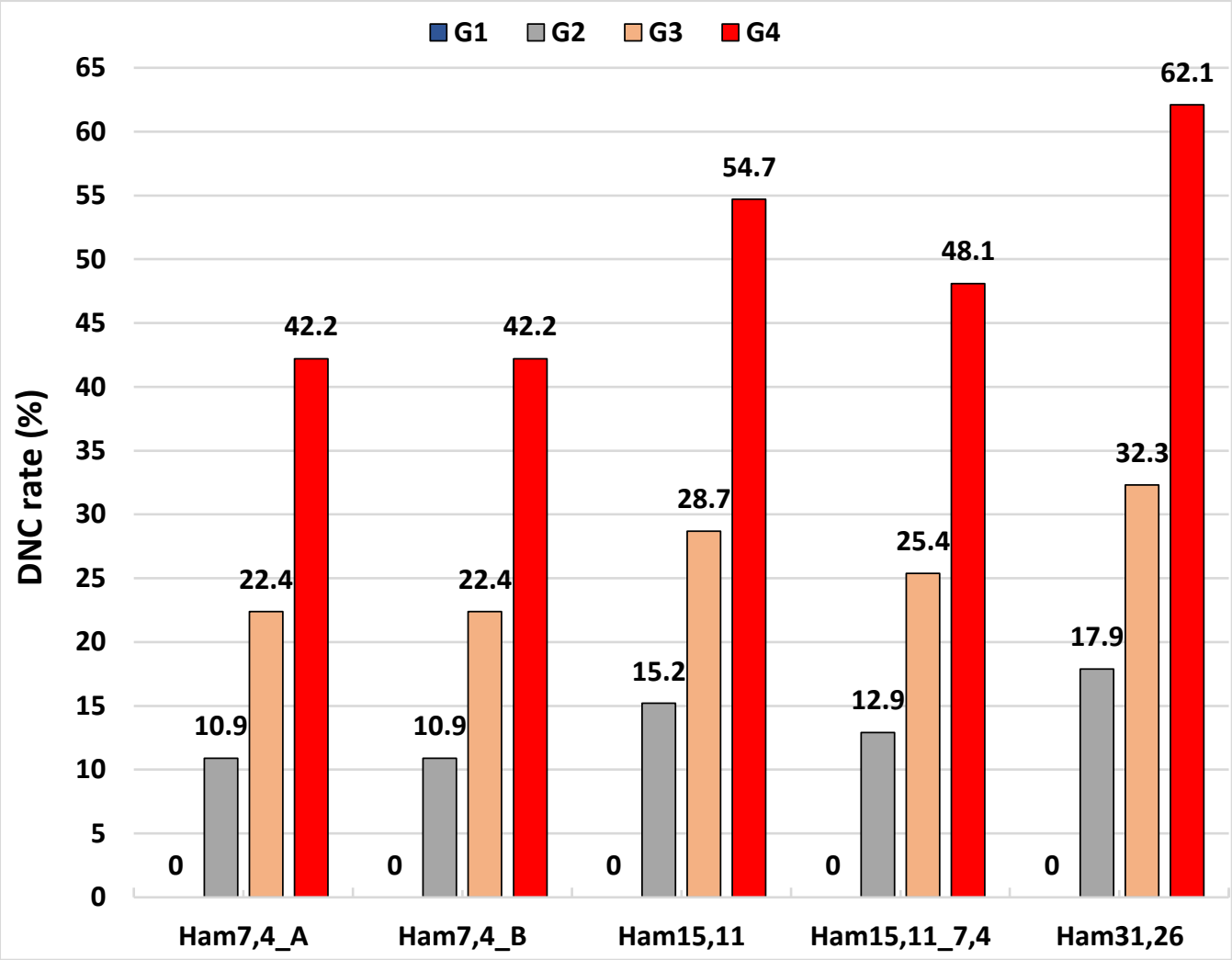

Fig. 6 Simulation results regarding **DNC** rate for the five ECC configurations and four bit-flip scenarios: (G1) a single bit-flip; (G2) two bit-flips; (G3) three bit-flips; and (G4) four bit-flips.

Fig. 6 displays that as the number of bit-flips increases, the DNC rates increase because all evaluated codes can correct only one error and detect two. The $Ham_{7,4_A}$ and $Ham_{7,4_B}$ configurations have the same DNC rates for all bit-flip groups. This similarity shows that, in this case, the separation of the ECCs does not affect the detection rate. The $Ham_{15,11}$ configuration obtained higher DNC rates than the $Ham_{15,11_7,4}$ configuration, even though it has a lower redundancy; i.e., while the $Ham_{15,11}$ configuration has only 8 bits of redundancy, the $Ham_{15,11_7,4}$ configuration has 10 bits of redundancy. A determining factor for decreasing the DNC rates of $Ham_{15,11_7,4}$ is that this configuration has three uncovered bits, whereas the $Ham_{15,11}$ configuration has only two uncovered bits.

Fig. 7 shows the ND rate for all configurations and scenarios evaluated in the experimental results. When only one bit-flip occurs, $Ham_{31,26}$ has the lowest ND rate, 3.1%, whereas $Ham_{7,4_A}$ and $Ham_{7,4_B}$ are the configurations that have the highest ND rate for scenario G1. Regarding scenarios G2 and G3, the highest ND rates are attained with the $Ham_{7,4_B}$ configuration. $Ham_{7,4_A}$ and $Ham_{7,4_B}$ also have the highest ND rates concerning scenario G4, both configurations with 22.1%, while the $Ham_{31,26}$ achieved only an ND rate equal to 6.5%.

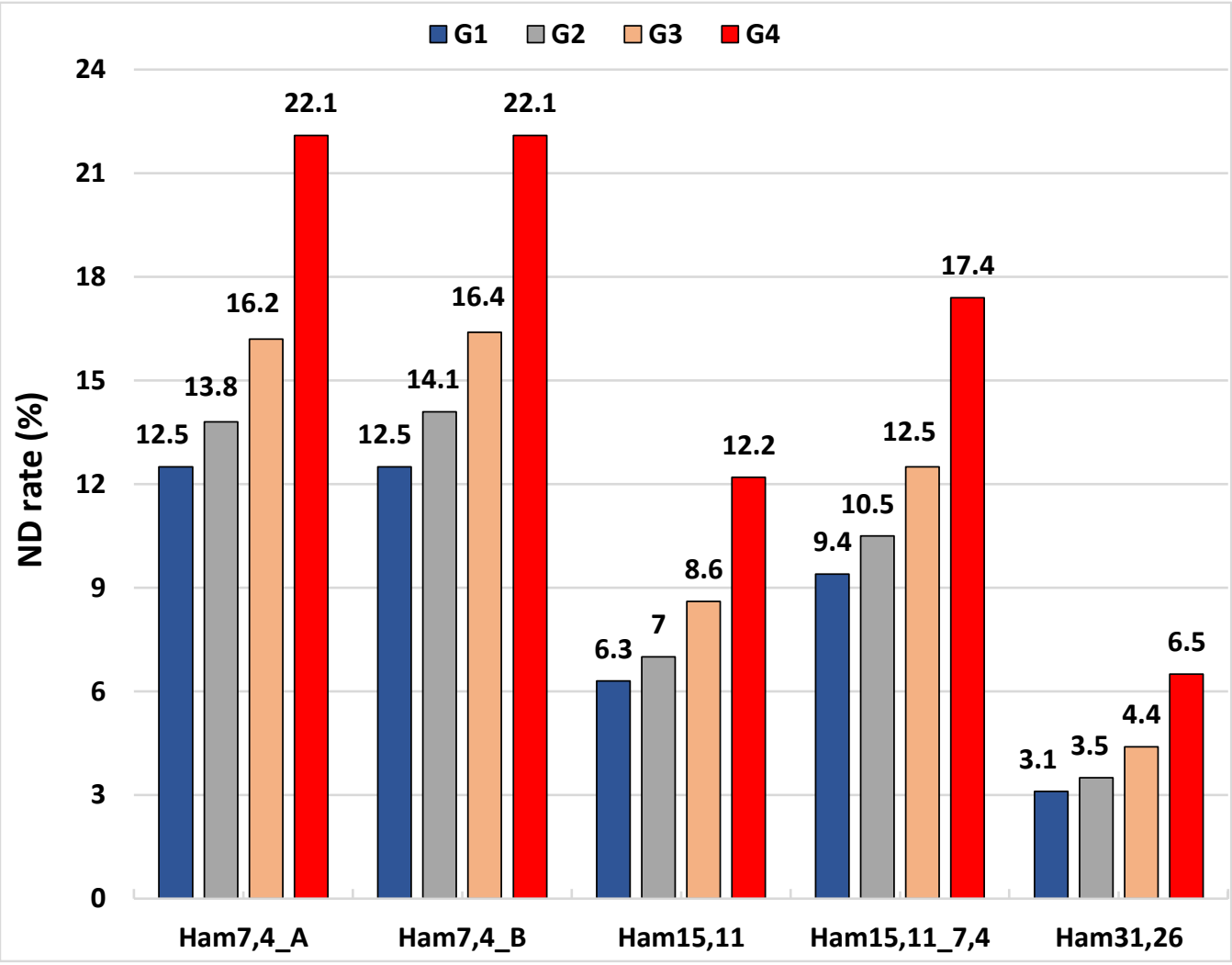

Fig. 7 Simulation results regarding **ND** rate for the five ECC configurations and four bit-flip scenarios: (G1) a single bit-flip; (G2) two bit-flips; (G3) three bit-flips; and (G4) four bit-flips.

## C. Results of Not Corrected Errors

The $Ham_{7,4_B}$ configuration has ND rates higher than the $Ham_{7,4_A}$ configuration, showing that the distance between ECCs does not reduce the number of undetected errors. For all scenarios evaluated, the $Ham_{15,11_7,4}$ configuration has ND rates higher than the $Ham_{15,11}$ configuration. Finally, the $Ham_{31,26}$ configuration again attained the best experimental results. All the ND rate results are directly related to the number of bits uncovered by the ECCs; i.e., the ND rate is directly proportional to the number of uncovered bits.

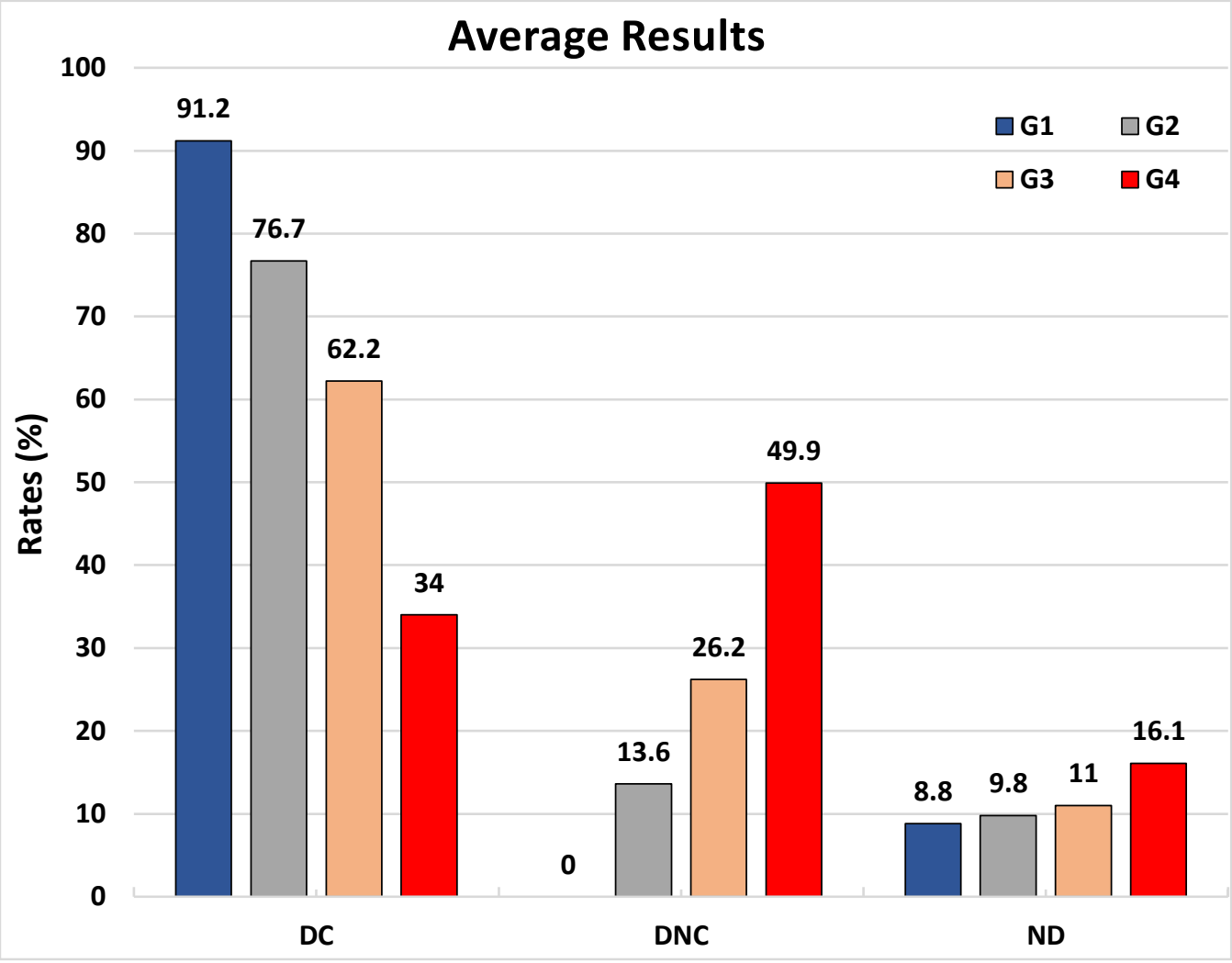

Fig. 8 Mean of the DC, DNC, and ND rates obtained with the simulations of the five ECC configurations evaluated in each of the four error scenarios.

Fig. 8 groups the mean value of all rates shown in Fig. 5 to Fig. 7. According to Rao et al. [26], more than 50% of errors in 45nm SRAM CMOS technology are only in a single bit. Scenario G1 shows that 91.2% of this error are corrected, and only 8.8% are not detected, on average. As the number of bit-flips increases, the corrected bit rate decreases, and the detection rates without correction and non-detection increase. For the case of the four bit-flip scenarios (G4), the detection and non-correction rate is higher than the correction rate. However, the sum of DC and NDC rates reaches almost 84% of the cases. This situation is less relevant for highly reliable systems since the higher interest is in detecting simple faults.

## D. Average Results of DC, DNC and ND, and Discussions

Fig. 9 shows the mean value of the correction and detection rates of the thirty-six error patterns. The correction rates are higher with single-error patterns in a line, for example, patterns 1, 3, 4, and 6. However, the correction rate falls sharply with double errors in the same line; for example, the DC rates of patterns 2 and 5 are 12.2% and 23%, respectively. The DC rates of patterns like 12, 13, 15, and 16, which have double errors on the same line but also have simple errors in other lines, are higher than the DC rates acquired with patterns 2 and 5 because they have simple errors that are always detected and corrected if they are in a region covered by an ECC.

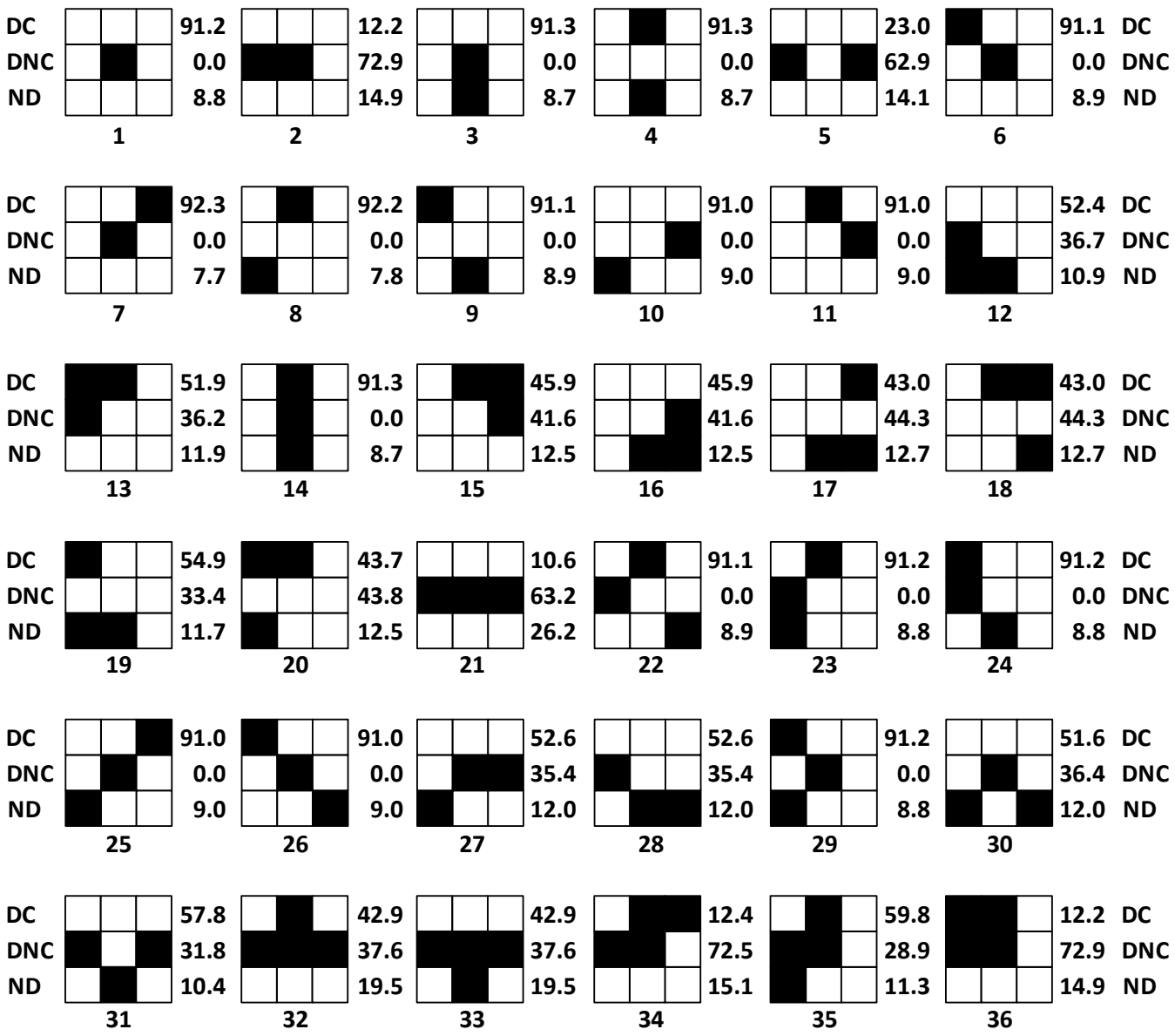

Fig. 9 Average results of the detection and correction rates of the error patterns used in the experiments. The DC, DNC, and ND values are on the right side of each error pattern, from top to bottom.

Error patterns 21, 34, and 36 have the lowest DC rates (i.e., 10.6%) that occur with error pattern 21 since it has three errors on the same line. Even if the triple error occurs in the interaction between two ECCs in the same line, there will always be an ECC that will have two bit-flips, not having the ability to correct them, only to detect them. Error patterns 34 and 36 also have low DC rates because they have two lines with double errors, causing a DC rate to be around 12%.

## E. Reliability Analysis

This work analyzes code reliability using [33] and [34]. We will assume the following statements that were also assumed in [34]: (i) transient faults occur with a Poisson distribution, and (ii) bit faults are statistically independent. According to these definitions, Equation 4 describes the fault correction in $F_c(t)$.

$$F_c(t) = \sum_{i=1}^{Ne} (P\{FC|iF\} \times P\{iF|MF\}) \quad (4)$$

$Ne$ is the maximum number of errors that can arise during time $t$, $FC$ is the number of corrected errors, $MF$ indicates if the memory faults, and $iF$ indicates that there are $i$ faults in the memory. Equation 5 describes the probability of having exact $i$ upsets in memory when memory is faulty.

$$P\{iF|MF\} = \frac{P\{iF\}}{P\{MF\}} \quad (5)$$

$P\{iF\}$ is given by

$$P\{iF\} = \binom{n}{i} (1 - e^{-\lambda t})^i e^{-\lambda(n-i)t} \quad (6)$$

In Equations 6 and 7, $n$ is the number of bits in the codeword, $\lambda$ is the bit-fault rate, and $t$ is the time parameter. Equation 7 describes the memory failure probability.

$$P\{MF\} = 1 - e^{-\lambda nt} \quad (7)$$

$P\{FC|iF\}$ values are obtained in the previous section through the simulation results presented in [34]. Besides, memory reliability is the product of the reliability of all words and can be given by Equation 8. $M$ is the number of words in memory (this work uses $M = 50$). Additional information about the equations can be found in [31].

$$R(t) = \left(1 - P\{MF\} + \sum_{i=1}^{Ne} P\{iF\}P\{FC|iF\}\right)^M \quad (8)$$

Ham(31,26) reaches the highest reliability over the entire period analyzed. Ham(7,4)-A and Ham(7,4)-B have practically equal values because their correction rates are very similar. Concerning the other two ECC configurations, Ham(15,11) has better reliability than Ham(15,11)(7,4), even though it has a lower correction rate for G4 of Fig. 10.

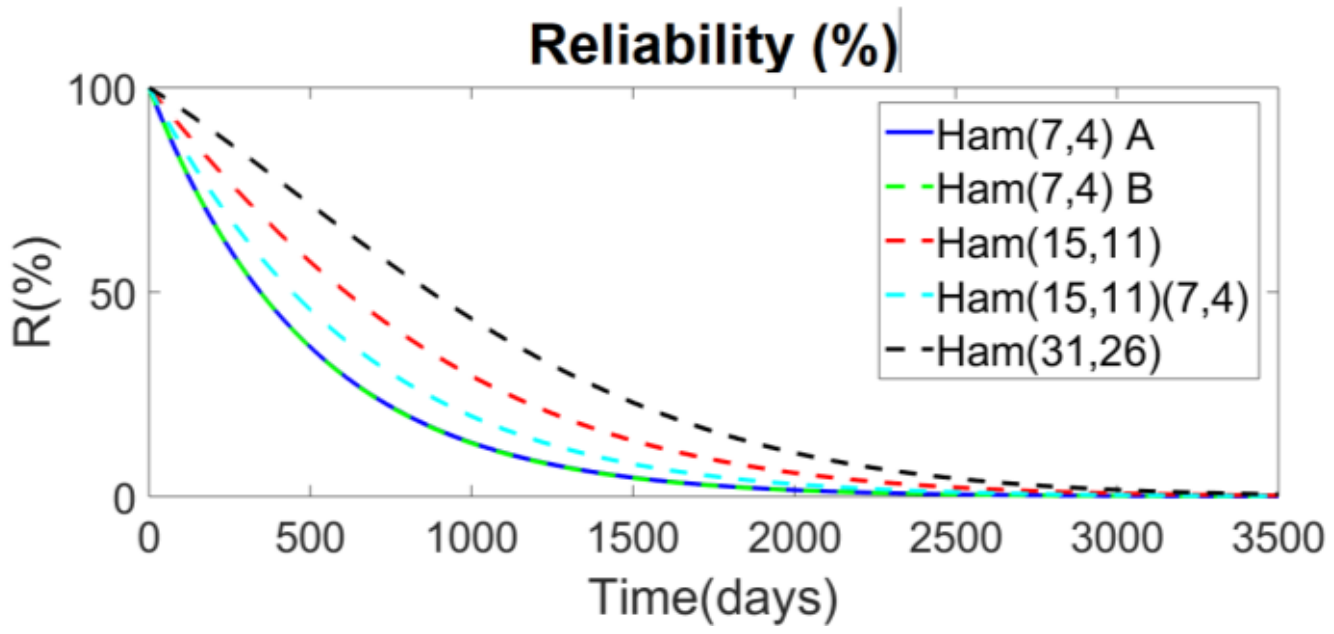

Fig. 10 Reliability of the ECCs presented as a function of time of a 32-bit memory with M = 50.

TABLE I
REDUNDANCY RATES FOR ALL ECC CONFIGURATIONS.

| Configuration | $tr$(%) |
|---|---|
| $Ham_{7,4_A}$ | 42.9 |
| $Ham_{7,4_B}$ | 42.9 |
| $Ham_{15,11}$ | 26.7 |
| $Ham_{15,11_7,4}$ | 34.5 |
| $Ham_{31,26}$ | 16.1 |

For time t=500, the worst reliability is Ham(7,4)-B with 36.47%, followed by Ham(7,4)-A with 36.66%. For Ham(15,11)(7,4), the value is 45.7%. The two largest reliabilities are 71.43% and 57.36%, respectively, for Ham(31,26) and Ham(15,11).

### F. *Redundancy Analysis*

Let $r$ be the number of redundancy bits, and $n$ be the number of codeword bits; then, Equation 9 presents the computation of the redundancy rate $tr$.

$$tr = \frac{r}{n} \quad (9)$$

Table I shows $tr$ for all five Hamming code configurations. The higher $tr$, the greater the weight of the redundancy bits concerning the entire codeword. In this way, the lower the $tr$, the lower the ECC cost. Consequently, besides the Ham(31,26) configuration having the highest DC rate results, it still has the lowest value of $tr$, making Ham(31,26) the most suitable configuration to be applied as ECC for the memory evaluated in the experiments.

## IV. CONCLUSION

This article explores the capabilities of five configurations with three types of Hamming codes in detecting and correcting bit-flips; the analysis was performed in an example of 8×32-bit memory. The experiments show that Ham(31,26) obtains the best results; In this organization, twenty-six of thirty-one bits are encoded in a 32-bit word, and only one bit remains uncovered. Ham(31,26) corrects almost 97% of simple errors, and the correction rates for double, triple, and quadruple errors are 78.7%, 63.4%, and 31.4%, respectively. Finally, undetected error rates for one-, two-, three- and four-bit errors are 3.1%, 3.5%, 4.4%, and 6.5%, respectively.

For all ECC configurations, the correction capacity decreases as the number of bit-flips increases. Another relevant point is that the configuration performance improves as the redundancy rate $tr$ decreases (i.e., fewer redundancy bits and more data bits in the 32-bit word). The main element for this improvement is the number of bits not covered by the code. For Hamming configurations with low $tr$, the number of bits not covered by the ECCs is smaller and increases as $tr$ increases.